\documentclass{aa}
\usepackage{graphicx}
\usepackage{times}
\usepackage{xspace}

\usepackage{savesym}
\usepackage{amsmath}
\savesymbol{iint}
\usepackage{txfonts}
\restoresymbol{AMS}{iint}

\usepackage{amssymb}
\usepackage{url}
\usepackage{natbib}

\newcommand{\sax}{\textsl{BeppoSAX}\xspace}
\newcommand{\xte}{\textsl{RXTE}\xspace}
\newcommand{\kevnxs}{ke\kern -0.09em V}
\newcommand{\kev}{\kevnxs\xspace}
\newcommand{\pca}{\textsl{PCA}\xspace}
\newcommand{\hexte}{\textsl{HEXTE}\xspace}
\newcommand{\asm}{\textsl{ASM}\xspace}
\newcommand{\qcm}{\ensuremath{\text{cm}^2}\xspace}
\newcommand{\ca}{\ensuremath{\sim}}

\newcommand{\degmark}{\ensuremath{^\circ}}
\def \nh {N${\rm _H}$}
\def \ergsec{\hbox{erg s$^{-1}$}}
\def \hcm {\hbox {\ifmmode $ atom cm$^{-2}\else atom cm$^{-2}$\fi}}
\def \arcmin {\hbox{$^\prime$}}

\def \chisq {$\chi ^{2}$}
\def \rchisq {$\chi_{\nu} ^{2}$}
\def\approxgt{\mathrel{\hbox{\rlap{\lower.55ex \hbox {$\sim$}}
        \kern-.3em \raise.4ex \hbox{$>$}}}}
\def\approxlt{\mathrel{\hbox{\rlap{\lower.55ex \hbox {$\sim$}}
        \kern-.3em \raise.4ex \hbox{$<$}}}}

\newcommand{\4}{4U2206$+$54\xspace}

\def \ergseccm{\hbox{erg s$^{-1}$} cm$^{-2}$}

\begin{document}

   \title{Evidence for a Neutron Star in the non-pulsating massive X-ray binary \4.}

   \author{J.M. Torrej\'{o}n
          \inst{1}
          \and
          I. Kreykenbohm
	  \inst{2,3}
 	  \and
	  A. Orr
	  \inst{4}
	  \and
	  L.Titarchuk
	  \inst{5}
	  \and
	  I.Negueruela
	  \inst{1}
}

   \offprints{J.M. Torrej\'{o}n (\texttt{jmt@disc.ua.es})} 
   
   \institute{ Departamento de F\'{\i}sica, Ingenier\'{\i}a de
     Sistemas y Teor\'{\i}a de la Se\~{n}al, Escuela Polit\'{e}nica
     Superior, Universidad de Alicante, Ap. 99, E-03080 Alicante,
     Spain \and Institut f\"ur Astronomie und Astrophysik --
     Astronomie, Sand 1, D-72076 T\"ubingen, Germany \and Integral
     Science Data Centre, Chemin d'Ecogia 16, 1290 Versoix,
     Switzerland \and ESA/ESTEC Research and Science Support Division,
     Noordwijk, The Netherlands \and US Naval Research Laboratory,
     Space Science Division, 4555 Overlook Av., Washington, USA}

   \date{Received November 25 2003; accepted April 30 2004}

   \abstract{We present an analysis of archival \xte and \sax data of
     the X-ray source \4 . For the first time, high energy data ($\ge
     30$\,\kev) 
     are analyzed for this source. The data are well described by
     comptonization models (\texttt{CompTT} and \texttt{BMC}) in which seed photons with temperatures
     between 1.1\,\kev and 1.5\,\kev are comptonized by a hot plasma
     at 50\,\kev thereby producing a hard tail which extends up to, at
     least, 100\,\kev. We offer a new method of identification of neutron star systems using a temperature - luminosity relation. If a
given X-ray source is characterized by a low bolometric luminosity and a
relatively high color
blackbody temperature ($>1$ \kev) it has necessarily to be a neutron
star rather than a black hole.
     From these arguments it is shown that the area of the soft
     photon source must be small ($r\approx 1$\,km) and that the
     the accretion disk, if present, must be truncated very far from
     the compact object. 
     Here we
     report on the possible existence of a cyclotron line around
     $30$\,\kev. The 
     presence of a neutron star in the system is strongly favored by the
     available data. 

   }

   \titlerunning{A Neutron Star in the HMXRB 4U2206+54}

   \maketitle
%

\section{Introduction}

High Mass X-ray binaries (HMXRBs) are often subdivided in two broad
categories depending on the type of the optical companion: Supergiant
systems and Be/X systems. The vast majority of these High Mass X-ray
Binaries harbour X-ray pulsars (Bildsten et al. 1997). Around 70\% of
all known X-ray pulsars are in Be/X-ray systems. These are believed to
be young neutron stars (NS) with relatively strong magnetic fields ($B\sim
10^{12}$ G). The X-ray spectra of these objects are usually described by powerlaws
modified at high energies by a cutoff around 7--10 \kev (Lewin et al. 1995).
 Some HMXRBs do not display X-ray
pulsations. From these, three show the typical characteristics of black hole
candidates, two of them being in the LMC and only one (Cyg X-1) in the
Galaxy.

There are, however, other HMXRBs in which pulsations have not been
detected, in spite of intensive searches, that do not display the 
characteristics of black hole candidates. The nature of their compact
objects is therefore unknown. One such system is \4.

\4 was first reported by Giacconi et al. (1972) using the
\textsl{Uhuru} satellite. Steiner et al. (1984) identified the optical
counterpart with the early type star BD$+53\degmark 2790$. This star
showed H$\alpha$ in emission while the photometric colours where
consistent with a B0 star and, consequently, the X-ray source was
proposed as a Be/X-ray binary. It is a persistent X-ray source, having
been detected by all satellites that have observed it. 
The luminosity is low ($L_\text{X}\sim 10^{35}$\,\ergsec) and fairly
constant in the long term but shows flaring with changes of an
order of magnitude in short timescales.
Fits to \textsl{EXOSAT} (Saraswat \& Apparao 1992) and
\textsl{RXTE} (Negueruela \& Reig 2001, NR01 hereinafter) data 
favored powerlaw models. Using \textsl{PCA} data on board
\textsl{RXTE} Corbet \& Peele (2001, CP01) did not find any evidence of
pulsations. Instead they found a modulation in the \textsl{ASM} X-ray lightcurve
with a period of $\sim$9.6\ d, which can be interpreted as the binary
period.

On the other hand NR01 concluded that the
spectral classification of the optical companion is far from being
well established. The optical spectrum does not belong to any specific
spectral type and is rather peculiar. These authors conclude that
BD$+53\degmark 2790$ is not a classical Be star but that the most
probable scenario is a peculiar active O9p star of
moderate luminosity (III to V) orbited by a compact object which
accretes matter from the stellar wind.

The nature of the compact object, in turn, is not clear. The presence
of a cutoff at $\sim 7$\,\kev in the X-ray spectrum (NR01, CP01)
favours a neutron star interpretation. However, the lack of pulsations
does not allow to confirm it unambiguously. The system displays
striking analogies with the microquasar RX J1826.2-1450$-$LS 5039
(optical counterpart O6.5V[(f)], low persistent luminosity and no
pulsations). In the case of \4  the
presence of a black hole would be consistent with the above
characteristics (especially the lack of pulsations). However, as we
will show in this paper, there is strong evidence for the presence of
a NS in the system.

So far, the spectral analysis has been limited to the 2--30\,\kev 
band. The spectra have been usually fit using  
powerlaws modified at high energies by cutoffs. 
In order to gain a more physical insight into the
properties of this mysterious system we present in this paper an
analysis of partially unpublished archival \xte data and unpublished \sax data of three different 
epochs.  \4 was observed in a very broad energy range from 0.5 to
200\,\kev by \sax and from 2.5 to 100\,\kev by \xte. For the first
time, data from the High Energy X-ray Timing Experiment (\hexte)
onboard \xte are analysed for this source.

\section{X-ray observations}

\begin{table}
\caption{Journal of the observation for the 1998 November 23 \sax
  observation. T$_\text{int}$ is the net integration time in ks. The
  net integration time of PDS is shorter by $\sim$50\% due to the
  high deadtime of the instrument. The net integration time of the
  LECS is $\sim$50\% shorter as well due to the fact that it observes
  only during night time. The beginning and the end of the observation
  is in MJD.} 
\begin{tabular}{lcccr}
\hline
\hline 
Detector   & c/s & Start & Stop &  T$_{\rm int}$ \\
\hline
LECS & 0.197$\pm$0.005 & 51140.710 &  51140.853 & 12.4\\
MECS & 0.404$\pm$0.004 & 51140.685 &  51140.828 & 33.4\\
PDS & 0.451$\pm$0.006 &  51140.686 &  51140.829 & 14.6\\
\hline
\end{tabular}
\label{tab:obs_times}
\end{table}

\begin{table}
\caption{\xte observed \4 in 1997 March and 2001 October. Provided are
  \xte 
  observation ID, the beginning and the end of the observation in
  MJD, and the resulting 
  on-source time after initial data screening and dead time 
  correction for the \pca and the \hexte in ksec. For details of the data
  extraction, see text.} 
\label{tab:xte_obs}
\begin{tabular}{lrrrr}
\hline
\hline 
ObsID & Start  & Stop  & \pca  & \hexte \\
\hline
20140-01-01-00 & 50518.176 & 50518.407 & 9.1 & 5.9\\
20140-01-01-01 & 50520.251 & 50520.365 & 4.5 & 3.1\\
\hline
60071-01-01 & 52194.010 & 52194.164 & 17.7 & 12.2 \\
60071-01-02 & 52196.268 & 52196.601 & 29.8 & 19.2 \\
60071-01-03 & 52198.834 & 52199.017 & 20.9 & 13.9 \\
60071-01-04 & 52201.339 & 52201.433 & 19.6 & 12.6 \\
\hline
Total & & & 101.6 & 66.9 \\
\hline
\end{tabular}
\end{table}

\subsection{BeppoSAX data}

\begin{figure*}
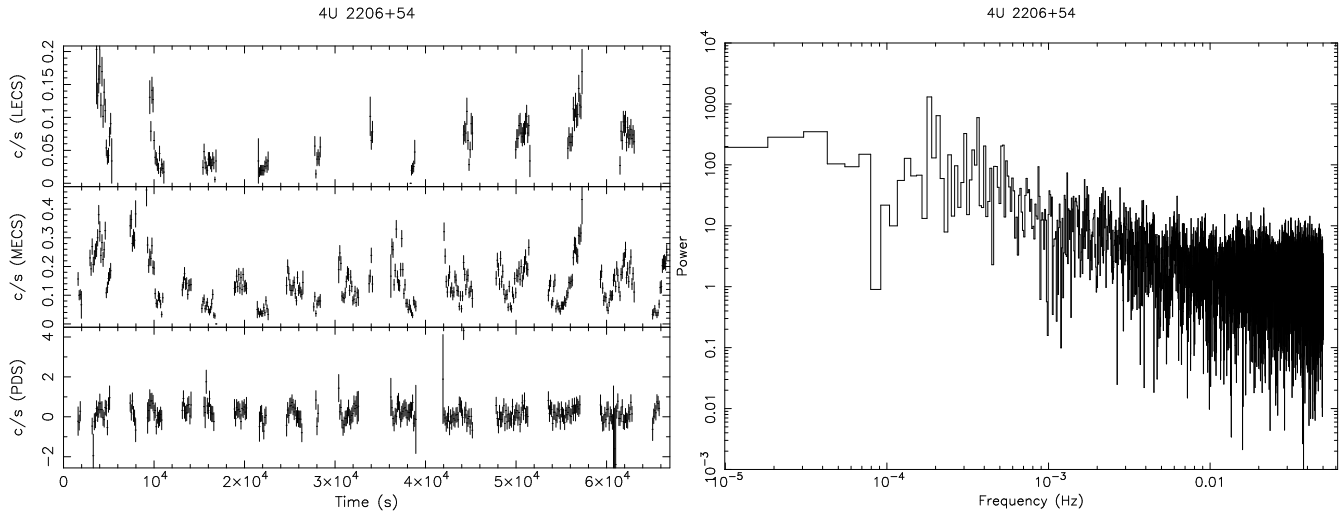

\includegraphics[height=\columnwidth, angle=-90]{0743fg1a.ps}
\includegraphics[height=\columnwidth, angle=-90]{0743fg1b.ps}
\caption{\textsl{Left panel:}0.1--10.0\,\kev LECS, 1.65--10\,\kev MECS
   and 15 - 200\ \kev PDS lightcurves . Data have been rebinned to 60\,s. \textsl{Right
   panel:} MECS power spectrum. In agreement with previous studies, no
   significant period is found. The same is true for the other instruments.
   } 
\label{lightcurve}
\end{figure*}

We present an observation made in 1998 November 23 using the 
Low-Energy Concentrator Spectrometer (LECS; 0.1--10\,\kev; Parmar et
al. \cite{p:97}), the Medium-Energy Concentrator Spectrometer (MECS;
1.8--10\,\kev; Boella et al. \cite{b:97}) and the Phoswich Detection
System (PDS; 15--300\,\kev; Frontera et al.  \cite{f:97}) on-board
\sax. 
All these instruments are coaligned and referred to as Narrow Field
Instruments (NFI).  The MECS consists of two (three until May 9, 1997)
grazing incidence telescopes with imaging gas scintillation
proportional counters in their focal planes. The LECS uses an
identical concentrator system as the MECS, but utilizes an ultra-thin
entrance window and a driftless configuration to extend the low-energy
response to 0.1\,\kev. The non-imaging PDS consists of four independent units arranged in
pairs each having a separate collimator. Each collimator was
alternatively rocked on-source and 210\arcmin\ off-source every 96\,s
during the observation. The High Pressure Gas Scintillation Proportional Counter on-board \sax
(HPGSPC) was not used in
this observation as the source was not bright enough for this
instrument. Table \ref{tab:obs_times} lists the \sax observation epoch and net 
exposure times for our source. The LECS instrument is able to observe only
  during night time. Therefore, the on-source time for LECS is usually
  a half that of MECS.

Good data were selected from intervals when the elevation angle above
the Earth's limb was $>4^{\circ}$ and when the instrument
configurations were nominal, using the SAXDAS 2.0.0 data analysis
package. In order to produce the spectra and lightcurves, LECS and MECS data
were extracted centered on the position of the source using the
standard radii of 8\arcmin\ and 4\arcmin, respectively. Background subtraction for the PDS was performed in the standard way
using data obtained during intervals when the collimators were offset
from the source. Background subtraction for the imaging instruments
(LECS and MECS) was performed using blank sky fields provided by the
\sax Science Data Center, using the same region of the detector as the
source. 

The LECS and MECS spectra were rebinned to oversample the full width
half maximum of the energy resolution by a factor 3 and to have
additionally a minimum of 20 counts per bin to allow use of the
$\chi^2$ statistic. Data were selected in the energy ranges
0.1--10.0\,\kev (LECS), 1.65--10\,\kev (MECS) and 15--200\,\kev (PDS)
where the instrument responses are well determined and sufficient
counts are obtained.  This gives background-subtracted count rates of
0.197, 0.404, and 0.451 ~s$^{-1}$ for the LECS, MECS and PDS,
respectively. 

All uncertainties quoted are on a 90\% confidence level for one
parameter of interest unless otherwise specified.

\subsection{\xte data}

We used archival \xte data of \4 for this analysis. \xte observed \4
in AO2 from 1997 March 11 to 1997 March 12 and four times between 2001
October 12 and 2001 October 20. The observations are evenly 
distributed over the possible orbit. The 1997 observations were quite
short and resulted in less then 15\,ksec onsource time in the \pca.
The 2001 observation was substantially longer and resulted in almost
90\,ksec usable onsource time in the \pca. The \hexte exposure times
are substantially shorter due to the rocking of the two clusters and
the deadtime of the instrument (see below). A detailed overview of the
observations is given in Table~\ref{tab:xte_obs}.

\xte consists of three instruments (for a complete description
of the satellite, see Bradt et al. 1993): the Proportional Counter Array
(\pca Jahoda et al 1996), the High Energy X-ray Timing Experiment
(\hexte Rothschild et al 1998), and the All Sky Monitor
(\asm Levine et al. 1996).

The \pca consists of five co-aligned Xenon proportional counter units
with a total effective area of \ca6000\,\qcm and a nominal energy
range from 2\,\kev to over 60\,\kev. However, due to response problems
above \ca20\,\kev and the Xenon-K edge around \ca30\kev, we restricted
the use of the \pca to the energy range from 3\,\kev to 20\,\kev
(see also Kreykenbohm 2002). Since the instrument is pointing
continuously at the source, a background model is used for background
subtraction. We used the \textsl{Faint} model as is appropriate for
a dim source like \4 (see for a description Stark 1997). To
account for the uncertainties in the \pca response matrix, we used
systematic errors. 

On the other hand, we have estimated the galactic
diffuse background emission at the source position as measured by the
ROSAT PSPC instrument. Then we converted this count rate into fluxes
with the PIMMS tool provided by HEASARC. The predicted 3 - 10\ \kev
flux is less than $10^{-14}$\ \ergseccm\ and does not contaminate the
source spectrum.

The \hexte consists of two clusters of four NaI(Tl)/CsI(Na) Phoswich
scintillation detectors with a total net detector area of 1600\,\qcm.
However, early in the mission, an electronics failure left detector
three in cluster B unusable for spectroscopy. These detectors are
sensitive from \ca15\,\kev to over 200\,\kev; however, response
matrix, instrument background and source count rate, limit the energy
range from 18 to 100\,\kev in the case of \4. Background subtraction is done in
the \hexte by source-background rocking of the two clusters every
32\,s.

To improve the statistical significance of the data, we added the data
of both \hexte clusters (using a 1:0.75 weighted response matrix to
account for the loss of one detector). We further improved the
significance of the data by binning several channels together.

\subsubsection{Lightcurves}

In Fig.\ref{lightcurve} the \sax LECS, MECS and PDS lightcurves rebinned to
intervals of 60\,s are shown. Variations of a factor of 3 can be seen
on timescales of \ca30\,min. A search for pulsations using Scargle
periodgram and Epoch-folding techniques (see, e.g., Leahy et al. 1983)
failed to result in any significant peaks (see Fig.\ref{lightcurve},
right panel) in agreement with previous works (NR01, CP01).

The lightcurve clearly presents flaring activity with periods of
relative quiescence. The erratic flaring seems otherwise to be typical
of wind accretion, as observed in other HMXRBs like Vela X-1
(Kreykenbohm et al. 2002, Kreykenbohm et al. 1999).

\subsubsection{Spectrum}

\begin{figure*}
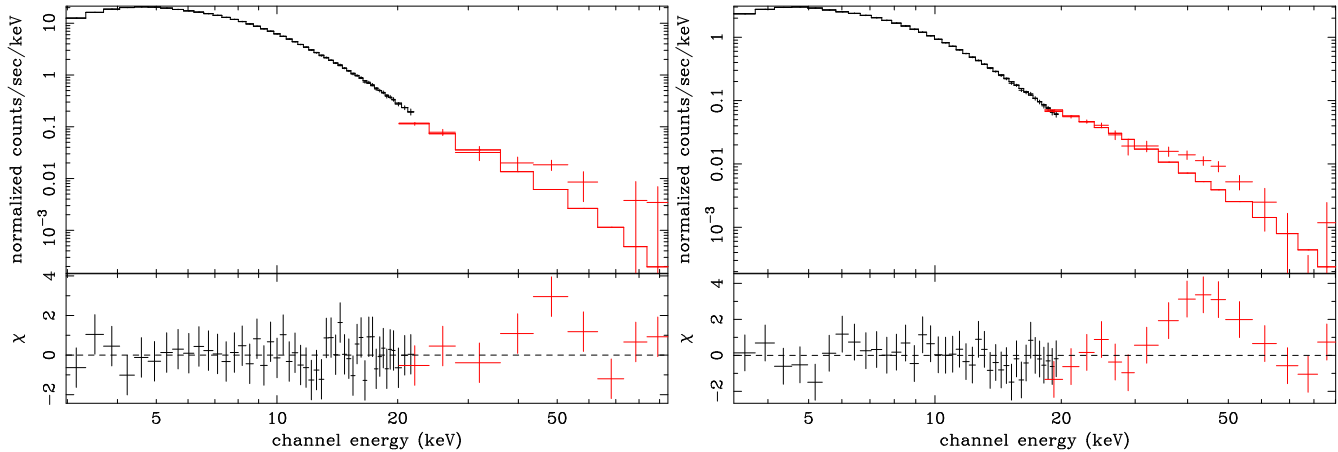

\includegraphics[height=\columnwidth, angle=-90]{0743fg2a.ps}
\includegraphics[height=\columnwidth, angle=-90]{0743fg2b.ps}
\caption{\textsl{Left panel:} XTE-1997 spectrum fit with a high
  energy cutoff powerlaw. The parameters are compatible with those
  obtained by NR(01). \textsl{Right panel:}
  XTE-2001 data with fit with the same model. As can be seen, the
  spectra is well described by the model between 3 and 30 keV. Beyond
  that point however, the residuals present a large bump. The same is
  seen in the SAX-1998 data.}
\label{highecutpo_models}
\end{figure*}

The overall spectrum of \4 was investigated by simultaneously fitting
data from the \sax NFI as well as the \xte \pca and \hexte data using
the \textsc{xspec} package (Arnaud \cite{a:96} v 11.2.00).

As can be seen in Table~\ref{tab:obs_times}, the on-source time is
very different for the three SAX instruments. Strictly speaking,
the MECS is sampling more epochs than LECS, while the source proved to
be very variable during the observation. However, the error bars in
LECS data are much larger than those of MECS so that this lack of
simultaneity has a very small incidence on the fit statistics. As a
matter of fact, the parameters of the fit are not too affected if we
reject the LECS data completely. If we reject all the MECS data not
strictly simultaneous with LECS then we obtain larger uncertainties in
the fit due to the lack of photons. We prefer, therefore, to use 
all available data.

The photoelectric absorption cross sections of Morrison \& McCammon
\cite{mo:83} and the solar abundances of Anders \& Grevesse \cite{a:89} were used throughout the analysis. 

Since much of the work has been done so far fitting modified powerlaws
up to 30 \kev the first thing we test is whether these models
fit properly the spectra if the high energy data is taken into 
account. In Table~\ref{powermodels} we quote the \chisq\ and degrees
of freedom for several models.

In Fig.~\ref{highecutpo_models} we present the XTE-1997 and
  XTE-2001 3-90\ \kev data fit with a high energy cutoff powerlaw. As can be
seen, the model describes correctly the data between 3 and 30 keV,
the spectral ranges used in the past. However, the model is unable to
describe properly the hard energy tail beyond 30 keV. In
the SAX-1998 data, the residuals present exactly the same behaviour
  and, in this case, do not describe properly even the low energy bins.  

\begin{table}
\caption{Goodness of fit for additive models during the three
  observations. The \rchisq\ is shown as well as the degrees of freedom
  (in parentheses). The name of the models are written in
  \textsc{xspec} terminology. All models include photoelectric absorption.}
  
\label{powermodels}
\begin{tabular}{lrrr}
\hline
\hline 
Model & XTE-1997  & SAX-1998  & XTE-2001 \\
\hline

po                 & 17(54) & 2.11(115) & 8.45(51)\\
highecut(po)       & 0.71(52)  & 1.32(113) & 1.27(49)\\
\hline
po + bb            & 0.94(52)  & 1.24(117) & 0.96(49)\\
po + diskbb        & 0.98(52)  & 1.26(117) & 1.31(49)\\
\hline
bremss             & 3.05(54) & 1.52(115) & 2.59(51)\\
raymond            & 17.83(54) & 1.87(115) & 10.65(51)\\
\hline 
\end{tabular}
\end{table}

\begin{table}
\caption{Parameters for the models of Table\
  \ref{powermodels} with lower \rchisq\ (second and third models)
  during the three pointings. The H column is given in units of
  $10^{22} \rm cm^{-2})$. Energies and temperatures are given in keV.}
  
\label{param_po_models}
\begin{tabular}{llccc}
\hline
\hline 
Model & Param & XTE-1997  & SAX-1998  & XTE-2001 \\
\hline

highecut(po) & $N_{\rm H}$ & 4.5$\pm 0.4$ &
1.1$\pm 0.3$ & 4.6$\pm0.1$\\

& E$_{cut}$ & 7.6$\pm 0.4$ & 7.8$\pm 0.5$ & 4.3$\pm 0.3$ \\
& E$_{fold}$ & 16.3$\pm 1.2$ & 11$\pm 3$ & 20$\pm 2$ \\
& $\Gamma$ & 1.6$\pm 0.1$ & 1.0$\pm 0.2$ & 1.6$\pm 0.1$\\
& & & & \\

po + bb  & $N_{\rm H}$ & 4.4$\pm 0.4$ & 1.0$\pm
0.3$ & 4.8$\pm 0.5$ \\  
& kT & 2.16$\pm 0.08$ & 1.74$\pm 0.11$ & 2.4$\pm 0.1$ \\
& $\Gamma$ & 1.96$\pm 0.04$ & 1.53$\pm 0.14$ & 2.0$\pm 0.1$\\
\hline 
\end{tabular}
\end{table}

\begin{figure*}
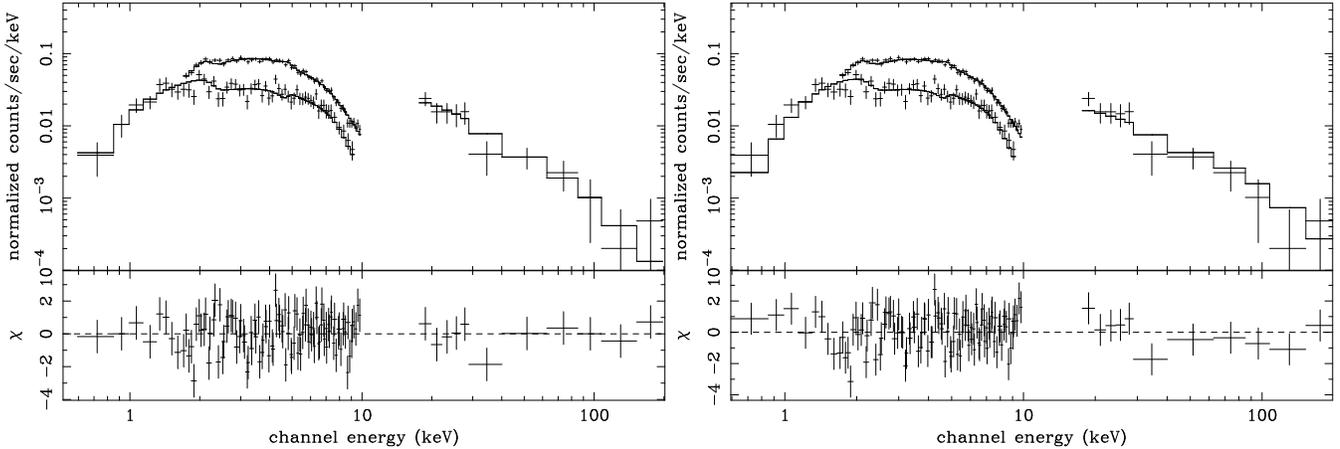

\includegraphics[height=\columnwidth, angle=-90]{0743fg3a.ps}
\includegraphics[height=\columnwidth, angle=-90]{0743fg3b.ps}
\caption{\textsl{Left panel:} SAX-1998 spectrum and folded BMC model. The model is only
  modified at low energies by photoelectric absorption. \textsl{Right panel:}
  Same data with a powerlaw plus blackbody model. }
\label{sax_bmc}
\end{figure*}

The analysis yields the following conclusions:

\begin{itemize}
  
\item The very presence of a hard tail is established for the
  first time, using high energy data from the \textsl{HEXTE} and
  \textsl{PDS} instruments. This hard tail extends at least to 100
  \kev where the signal to noise ratio starts to be poor.
  
\item Simple models including an absorbed powerlaw and an absorbed
  powerlaw with a high energy cutoff give poor fits when the hard tail
  beyond 30\ \kev is considered (see Fig.\ref{highecutpo_models})

\item Models including a soft blackbody or
  multicolored disk blackbody emission plus a hard powerlaw tail give
  good descriptions of XTE data but not for the SAX-1998 data
  (Fig.\ref{sax_bmc} right panel, specially at low energies). Note that
  the model with a multicolored blackbody disk results in a slightly
  worse fit (see discussion). 

\item Other models with combine thermal models and powerlaws, result in
poor fits. For XTE 2001 data, for instance, the \texttt{Bremss + po} model
results in a \rchisq\ of 2.5.

\item Models including only thermal components (Bremsstrahlung,
  Raymond, etc.) are unacceptable. 

\end{itemize}

Although in some instances the \texttt{powerlaw + bb} model could give an
acceptable fit, in others do not. In principle, this additive model is not
physically justified but it is a crude additive
approximation of the Comptonization model for which a blackbody component
and powerlaw component are the low energy and high energy asymptotics, 
respectively (see Sunyaev \& Titarchuk 1980, hereafter ST80). Thus we use
exact Comptonization models for model fitting to the data rather than any
approximation of it. In this work we shall deliberately
apply models which parameters can be more readily interpreted in
physical terms and have as few free parameters as possible. 

The best fits are obtained with 
the thermal comptonization model (compTT in \textsc{xspec})
(Titarchuk \cite{ti:94}, Hua \& Titarchuk \cite{ht:95}), with the addition of a blackbody component at low energies, 
and the Bulk Motion Comptonization model (Titarchuk et al. \cite{tmk:97}, Shrader \& Titarchuk \cite{st:98}), or BMC model.

This single
model reproduces rather well the data both at high and at low energies
where it must be only modified by photoelectric absorption. The best
fit parameters for these models are given in Table \ref{modelstab}.

\begin{table*}
\caption[]{Parameters for the comptonization models}
\begin{center}
\begin{tabular}{llccc}
\hline\noalign{\smallskip}
\hline\noalign{\smallskip}
Model & Parameter & XTE 1997 & SAX 1998 & XTE 2001\\
\noalign{\smallskip\hrule\smallskip}
 & \\
compTT + bb & N$_{\rm H}(10^{22} \rm cm^{-2})$& 2.04$\pm$0.12 & 0.30 $\pm$ 0.15 & 4.22$\pm$0.16\\
& T$_{\rm bb}$ & 0.76$\pm$0.04 & 0.73$\pm$0.05& 0.65$\pm$ 0.04\\
& kT$_{0}$ & 1.50$\pm$0.06 & 1.26$\pm$0.07 &1.46$\pm$0.06\\
& kT$_{e}$ & 47.79$\pm$0.06 & 47.48$\pm$0.06 & 46.46$\pm$0.06\\
& $\tau_{\rm p}$ & 0.50$\pm$0.3 & 0.70$\pm$0.3 & 0.46$\pm$0.3\\
& F$_{\rm 2-10~keV}^{(a)}$ & 26.8 & 4.1 & 12.6\\
& F$_{bol}^{(c)}$ & 67.0 & 5.4 & 16.5 \\ 
& CR$^{(b)}$ & 0.28 & 0.43 & 0.34 \\
& $\chi^{2}_{\nu}$(dof)& 0.66 (49) & 1.16 (114) & 0.62 (47)\\
 & \\
BMC & N$_{\rm H}(10^{22} \rm cm^{-2})$& 0.2$\pm$ 0.1 & 0.30 $\pm$ 0.15 & 0.2$\pm$0.1 \\
& $kT_{col}$~(keV) & 1.38$\pm$0.09 & 1.09$\pm$ 0.08 & 1.38$\pm$0.09  \\
& $\alpha$ & 1.4$\pm$0.23 & 0.99$\pm$ 0.18 & 1.26$\pm$0.15 \\
& $f$ & $\gg$1 & $\gg$1 & $\gg$1 \\
& $A_{\rm N}(\times 10^{-4})$ & 35.3$\pm$ 0.3 & 5.5$\pm$ 0.2 &15.7$\pm$ 0.2 \\
& F$_{\rm 2-10~keV}^{(a)}$ & 24.5 & 4.1 & 9.9\\
& CR$^{(b)}$ & 0.40 & 0.72 & 0.47 \\
& $\chi^{2}_{\nu}$(dof) & 0.71 (51) & 1.14 (116) & 0.88 (49)\\
\noalign{\smallskip\hrule\smallskip}
\end{tabular}
\\
$^{(a)}$ unabsorbed flux in units of 10$^{-11}$ \ergseccm\\
$^{(b)}$ computed as F$_{20-60\rm \kev}$/F$_{3-20\rm \kev}$\\
$^{(c)}$ unabsorbed flux in units of 10$^{-11}$ \ergseccm. The
spectral range \\
is 2-100\ \kev for RXTE data and 
0.5--100\ \kev for BeppoSAX data
\end{center}
\label{modelstab}
\end{table*}

Assuming a distance of $\sim $ 3 kpc (NR01), the luminosity of the
source was 7.22$\times 10^{35}$ \ergsec (XTE-1997, 'high state'),
5.8$\times 10^{34}$\,\ergsec (SAX-1998, 'low state') and 1.78$\times
10^{35}$ \ergsec (XTE-2001, 'intermediate state') respectively.
Consistently with earlier observations, no iron line around 6.4\,\kev
is detected.

\section{Discussion}

\begin{figure}
  \includegraphics[height=\columnwidth, angle=-90]{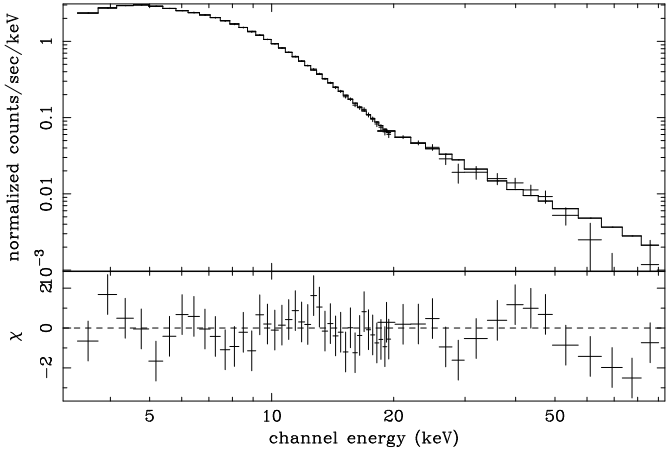}
  \caption{RXTE-2001 data fit with the BMC model. The model parameters
  are given in Table\ \ref{modelstab}.}
  \label{bmc_xte}
\end{figure}

As we have seen, the cutoff powerlaw is no longer a satisfactory
description when data beyond 30\,\kev are taken into account. Indeed the
spectra show clearly the presence of a hard tail which extends to the
100\,\kev zone. This hard tail can be described by a powerlaw  or by
comptonization models. 

In contrast, models including only thermal components can be discarded
(see Table \ref{powermodels}). This fact, along with the presence of
the hard tail, clearly argue against the accretion onto a WD
or the interaction between two normal stars as the origin of the X-ray
emission (a possibility put forward by NR01, owing to the unusual
optical spectrum). Indeed, colliding wind systems present spectra
which are rather soft with plasma at temperatures of $kT\sim 1$ keV
(Cooke, Fabian \& Pringle \cite{cfp:78}, Pollock \cite{po:87}). In
order to produce measurable X-ray emission the very strong stellar
winds found in WR stars are required. However the optical counterpart
of \4 is a late O~star of moderate luminosity (Class III or V).
According to NR01, the putative companion could be a B~star.
Such a system can not produce the hard X-ray emission observed in this
system. The presence of a BH or a NS is, therefore, required.

First we analyze the parameters deduced from the 'standard' thermal
comptonization plus blackbody emission model (\texttt{comptt
  + bb} in Table~\ref{modelstab}). Under this interpretation we first
note that the temperature of the injected soft photons ($kT_{0}$) is
rather high while the luminosity of the source is very small. In order
to reconcile these two facts, a small emission area must be invoked.
We therefore compute the radius of the Wien soft photon source by 
equating the bolometric luminosity of the soft photon source with that
of a black body of area $\pi R_{W}^2$ (In't Zand et al. 1999): 
 
\begin{equation}
R_{W}=0.6d_{\text{kpc}}\sqrt{\frac{f^{\text{bol}}_{-10}}{1+y}}(kT_{0}/1\
\rm \kev)^{-2}
[\text{km}]   
\end{equation}

where $y$ is the comptonization parameter
$y=4kT_{e}\tau^{2}/m_{e}c^{2}$ which gives the relative gain of energy
from the inverse Compton scattering. $f^{\text{bol}}_{-10}$ is the flux of the
comptonized component of the model, corrected for absorption, in units
of $10^{-10}$ \ergseccm. This is
obtained by integrating the spectrum. For the parameters given in Table
\ref{modelstab} we obtain $R_{W}\simeq 1.98$ km (XTE 1997), $R_{W}\simeq
0.88$ km (SAX 1998) and $R_{W}\simeq 0.96$ km (XTE 2001). 

On the other hand, we note that the temperature of the Compton cloud is
very high ($kT_{e}\sim$ 50~\kev), of the order of those found in BHCs.
The source is harder (higher color ratio) when the luminosity is lower
(see table \ref{modelstab}). However, these 'state transitions',
typical for BHCs, have been observed also in some pulsars specially in
low regimes.

The best single component model fit is achieved by applying the Bulk
Motion Comptonization model. This model reproduces rather well the
whole spectrum, without the addition of other components, both at high
and at low energies.

The BMC model (Titarchuk et al. \cite{tmk:97}, Shrader \& Titarchuk \cite{st:98}) is a general model for comptonization of soft photons
which uses the Green's (spread) functions for the treatment of
upscattering and which takes the form of a broken powerlaw. This
formalism is valid for any kind of comptonization (bulk comptonization
in first order $(v/c)$, thermal comptonization in second order
$(v/c)^2$) and remains valid up to photon energies comparable to the
mean plasma energy ($m_{e}c^2\sim 511~{\rm \kev}$ in the case of bulk
motion). This model has been applied successfully to BH
systems. 
The origin of the soft component is the innermost disk zone where the
gravitational energy of matter is released due to viscous dissipation
as well as geometric compression of the matter.  By itself, however,
this is not a proof for the presence of a BH in the system since it can
be applied also to accretion onto Neutron Stars. In this case the
source of soft photons is the surface of the compact object. Either
the disk or the surface (or both) emit a soft black body like spectrum
with a characteristic color temperature $T_{\rm col}$. The comptonizing
region (a compton cloud or a boundary layer) must cover effectively this
zone (i.e. the innermost region of the disk or the spot over the
surface) in order to be exposed to a high fraction of the seed photons.

\begin{figure}
\includegraphics[height=\columnwidth, width=\columnwidth, angle=0]{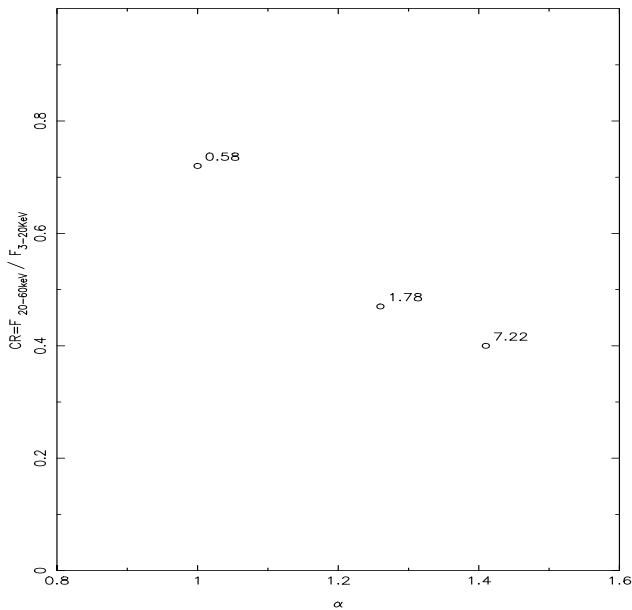}
\caption{Color ratio vs. $\alpha$ spectral index. The numbers
  represent the luminosity of the source in units of $10^{35}$
  \ergsec. Note that the source is softer when brighter.}

\label{CR_vs_alpha}
\end{figure}

One of the three free parameters of BMC model is a power law spectral
index $\alpha$ related to the Comptonization efficiency. When $\alpha$ is
smaller the efficiency is higher (for details see ST80). A value close
to 1 (vg. in SAX-1998 data) indicates
that the source is undergoing a phase transition from the low-hard to
high-soft state. These transitions can be
caused by the redistribution of mass accretion rates between Keplerian
(disk) and sub-Keplerian components or by an increase of the optical depth
$\tau_0$ for gravitational energy release at the shock (BH case) or at the
surface (NS case).  These spectral transition models  were studied and
discussed in detail by Zel'dovich \& Shakura (1969), for the NS case,
Chakrabarti \& Titarchuk (1995) for the BH case and Titarchuk \& Fiorito
(2004), for the BH and NS cases.
 In Figure \ref{CR_vs_alpha} we show the strong
correlation between the spectral index $\alpha$ and the color ratio
(CR). This plot shows that the source is softer when it is brighter
(because the upscattering is less efficient when luminosity
increases). That is to say, the
comptonization parameter decreases when luminosity (pressumably the
mass accretion rate) increases (Titarchuk \& Fiorito, 2004).  

The large $f$ value ($\gg$1), which is the ratio of the number of
photons multiply scattered in the converging inflow to the number of
photons in the thermal component, indicates that the soft spectrum has
been fully comptonized and that, as a matter of fact, we do not see
the source of soft photons at all. The Compton cloud completely
obscures the BMC region, namely, the innermost part of the accretion
disk or the surface of the NS.

Again, to reconcile the high color temperature of the soft emitting
region $T_{\rm col}\sim 1.1-1.3$~\kev with the low intrinsic
luminosity ($L_{\rm X}\sim 10^{35}\ergsec$) a small emission area must
be invoked.  Indeed, we can make use of the comptonization enhancement
factor $L/L_{0}$ where $L_{0}$ is the soft photon source luminosity
and $L$ is the resulting comptonized luminosity (Titarchuk et al. \cite{tmk:97}, eq.  37). For $\alpha\sim 1$ this is equivalent to
$\ln(1/x_{0})$ where $x_{0}=2.8kT_{\rm col}/kT_{e}$. Using the \sax\ 
data ($\alpha\sim 1$) from Table \ref{modelstab} and assuming a
Compton cloud of $kT_{e}\sim 50$ \kev (deduced from the thermal
comptonization model) we can derive the luminosity of the soft photon
source (which we do not see) and gives $L_{0}=1.6\times10^{34}$\ergsec.
Such a low luminosity cannot be produced in a inner accretion disk
of this temperature ($T_{\rm col}$). Borozdin et al. \cite{bo:99} (see Eq. 9 and Fig. 9 in that reference) show that a disk of temperature $\sim$ 1\ \kev has necessarily a relatively high luminosity of the order of $10^{36}-10^{37}$\ \ergsec that contradicts the observable luminosity of the soft component. Assuming that this luminosity is radiated by a
blackbody with an area $\pi R_{\rm W}^{2}$ and temperature $T_{\rm
  col}$ we can deduce (following the same argumentation to deduce
equation 1) that $R_{\rm W}\approx 0.6\sqrt{L^{34}_{0}}(kT_{col}/1\
\rm \kev)^{-2}$[km] where $L_{0}$ is in units of $10^{34}$ \ergsec. For the values given
above this yields 0.7 km. This is somewhat smaller than the values deduced
previously and stresses the need for a small emission area. Clearly,
it is not possible to have a contribution from a disk in this system
because the inner region of an accretion disk at these temperatures
would produce a much higher X-ray luminosity. Since the radius of the soft photon region is $\le 2$\ km, we conclude that the only viable candidate is a polar cap on a NS surface.

The BMC model implies an absorption column which is compatible with
the expected interstellar value (\nh $\sim0.3\times 10^{22}
\rm{cm}^{-2}$, using the relation of Bohlin et al. \cite{bo:78} and
$E(B-V)\simeq 0.5$ deduced from optical data, NR01). This can be
explained if the resulting comptonized spectrum is not absorbed
further beyond the comptonizing material close to the X-ray source.
That is to say, the circumstellar environment (surrounding the optical
star) should be tenuous at the site of the compact object.  This is
consistent with the hypothesis that the compact object orbits a main
sequence 'classical' O star (i.e. with neither a strong stellar wind, as
in supergiants nor a circumstellar envelope as in Be type stars) and
also with the lack of any detectable fluorescence iron line produced
in a cold thick surrounding medium.

This rarefied medium also contributes to the low luminosity. Indeed
the mass loss ratio between a O9.5V and a O9.5I star is (De Jager et
al. 1988): 

\begin{equation}
\frac{-\dot{M_{\text{V}}}}{{-\dot{M_{\text{I}}}}}\simeq 10^{-2} 
\end{equation}

So, roughly speaking, we have a factor of 10$^{2}$ less material to
power the X-ray source in a Main Sequence star, like BD$+$53\degmark
2790 appears to be, than for a supergiant system like Vela X-1 and
thus also a luminosity lower by the same factor. 

 At present, none of the characteristics that can establish
unambiguously the nature of the compact object (pulsations, Type I
outbursts, mass function, etc) have been established for this system.
As discussed before, however, the system lacks an inner accretion
disk, and the emission area is of the order of 1 or 2 km. This is only
consistent with emission from a hot spot on a NS surface. Note that
this conclusion is based on two different models. 

\begin{figure}
\includegraphics[width=\columnwidth]{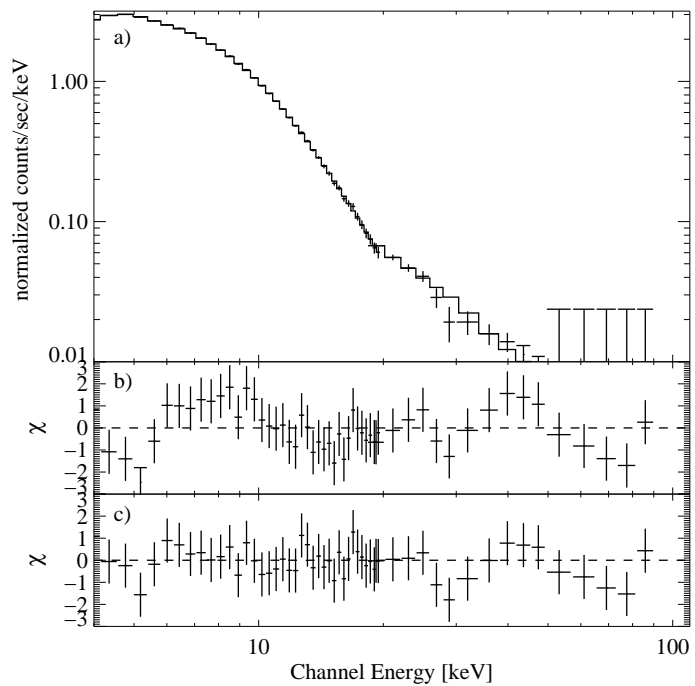}
\caption{\textbf{a} Data and folded model of \4 using the 2001-\xte
  observation. \textbf{b} shows the residuals when fitting the data
  with NPEX modified only by photo electric absorption. \textbf{c}
  same as b, but an additional soft black body component is
  included. }
\label{npex}
\end{figure}

\begin{figure*}
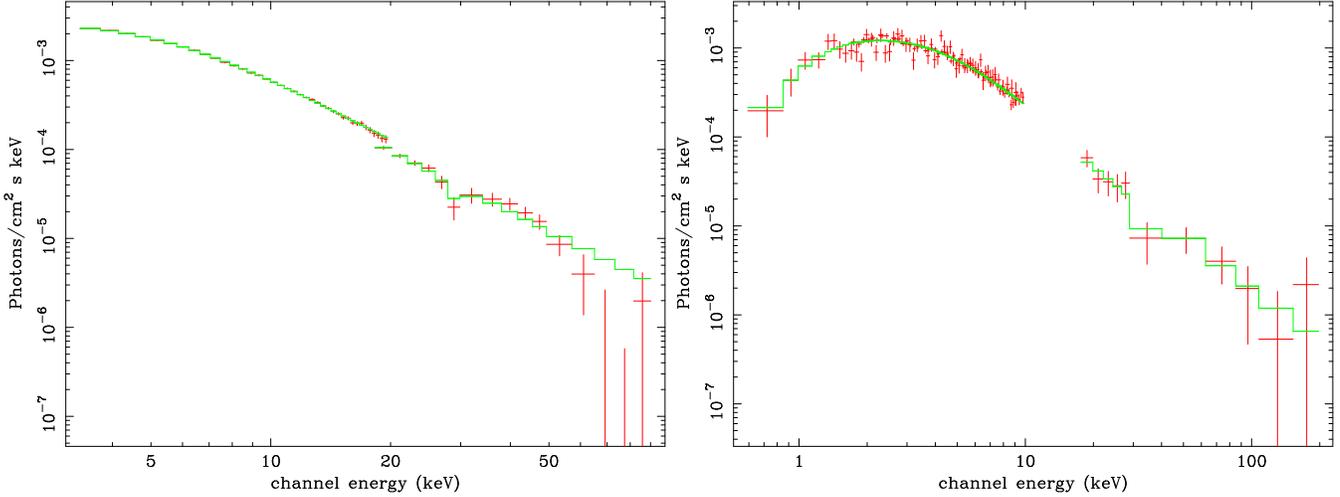

 \includegraphics[angle=-90, width=\columnwidth]{0743fg7a.ps}
 \includegraphics[angle=-90, width=\columnwidth]{0743fg7b.ps}
\caption{\textsl{Left panel:} Unfolded XTE-2001 spectrum of \4 showing the possible cyclotron
  line. We used the BMC model to describe the continuum. \textsl{Left
  panel:} Unfolded SAX-1998 spectrum of \4 showing the possible cyclotron
  line. The BMC model is used to describe the continuum. The line is
  at a slightly larger energy probably due to the coarse binning.}

\label{cyclabs_ufspec}
\end{figure*}

Other proofs, although circumstantial, also point to the NS
hypothesis. For instance the negative--positive exponential model
(NPEX Mihara et al. 1995) gives a comparably
satisfactory fit to the data. This model is commonly used to describe
the the spectra of pulsars
(Mihara et al. 1995, Makishimaet al. 1999, Kreykenbohm et al. 2002). 
The NPEX model has been successfully fitted to several neutron star
spectra like Vela~X-1 (Kreykenbohm 2002). For \4 the resulting fit
is acceptable (\rchisq = 1.2, see Fig.~\ref{npex}b) or
excellent (\rchisq = 0.6, see Fig.~\ref{npex}c), using an
additional low energy component like a black body. The 
parameters obtained are quite similar to those found in pulsar spectra.

\subsection{A cyclotron line at $\sim 30$\,keV?}

One final point deserves attention. In Figs.~\ref{sax_bmc} and 
\ref{bmc_xte}, an absorption dip around 30 \kev is observed. We were
first tempted to identify the dip as a cyclotron absorption. The
addition of a cyclotron line does, however, not improve the fit. 
Furthermore, it is only one energy bin in both spectra which is below
the continuum level, so an absorption feature cannot be claimed.
However, it is intriguing that the same feature can be seen in all the
spectra, i.e., spectra taken \emph{with different instruments and
  different epochs}.  Furthermore, although the \textsc{npex} model is
a completely different description of the spectrum, the unexplained
``dip'' at 30\,\kev is also present (see Fig.~\ref{npex}). 
Therefore, we are inclined to think that a very weak absorption
feature is really present in the data.

In order to obtain some information about the line we have first ignored
the energy bins from 25\,\kev to 32\,\kev. Then we fit the BMC 
model. Next we freeze these parameters, add a cyclotron absorption
feature (\texttt{cyclabs} in \textsc{xspec} terminology), notice again
energy bins from 25\,\kev to 32\,\kev, and refit. Finally we thaw all
parameters of the model and fit again. The continuum and line are
shown in figure \ref{cyclabs_ufspec}. For the XTE-2001 spectrum, the line is centered around
$E_{0}=29$ \kev, with a width of 1\,\kev and an optical depth
$\tau_\text{C}=0.6$.

Unfortunately the uncertainties are large and the analysis of the
background does not allow to reject the presence of calibration
effects so we can go no further on this issue with the available data. Further observations of the source at higher spectral resolution are required.

If we interpret this feature as a cyclotron resonant scattering
feature this would identify the source unambiguously as a neutron star
with a magnetic field of the order of $3.3\times10^{12}$\,G, taking into account the gravitational redshift. As a matter of fact we need a strong magnetic field if the material is
to be funneled to the small pole caps.

The Alfv\'{e}n
radius is inversely proportional to the accretion luminosity
and directly proportional to the magnetic field ($R_{M}\sim
m_{1}^{1/7}R_{6}^{-2/7}L_{37}^{-2/7}\mu_{30}^{4/7}$ where
$\mu=BR^{3}$, Frank et al. 2002).  \4 presents a combination of low
luminosity and thus, presumably, low accretion ratio, and (possibly) a
strong magnetic field. This radius could then be rather large and help to
truncate the accretion disk (if
any) very far from the compact object.

The lack of the inner accretion disk in the system can explain several
observational facts:

\begin{itemize}
\item the lack of a fluorescence iron line which would originate
in the cool parts of an irradiated disk 
\item the system presents no radio emission down to 0.04\ mJy
  (Rib\'{o} et al. priv.  comm.). Radio emission would be due to radio
  jets whose formation is strongly hampered without the inner disk
\item the lack of outbursts
  over the years: without an inner disk there
  is no large reservoir of matter which could undergo large scale
  perturbations; instead the source accretes directly from the thin
  steady stellar wind.
\end{itemize}

If the system has a strong magnetic field, why do not we observe
pulsations?. This could be explained by the very hot Compton Cloud
($kT_{e}\approx 50$\ \kev) of this system that can wipe out the pulsations (Titarchuk et al. 2002).  As a matter of fact, we see the radiation fully
comptonized while the source of soft photons (presumably the hot spot)
is completely obscured. Then it is not strange that we miss the
pulsations in this system. This leaves open the question, however, why
the compton cloud si so hot in this system and why we do not observe
pulsations in other similar systems. Of course, other possibility is that 
the rotational axis and the magnetic axis might not be offset (or very little offset). Further observations will be necessary to asses this issues.

\section{Conclusions}

\begin{enumerate}
     
\item The spectrum of \4 can be well described by comptonization
  models. These models show soft radiation with temperatures
  $kT\sim[1.1$--$1.5]$\,\kev comptonized by a hot plasma at
  $kT\sim$50\,\kev. This comptonization produces a hard tail visible
  up to, at least, 100\,\kev.
     
\item Application of the BMC Comptonization model to the data shows
that  the soft radiation is completely comptonized. Namely the  observer
does not see the direct blackbody radiation (presumably) from NS surface.
  
\item The temperature of the seed photons, along with the low
  luminosity point to a small emission area ($r\approx 1-2$ km radius)
  and contradicts the presence of an accretion disk.
  
\item The lack of a fluorescence iron line, the lack of radio emission
  and the lack of outbursts throughout the years also 
  argue against the presence of an accretion disk in this system.
  
\item Points 3 and 4 put stringent restrictions on the origin of the
  seed photons which must be emitted from a hot spot on a NS surface.
      
\item The X-ray lightcurve shows no evidence of any coherent
  pulsations at any explored energy range (0.5--60\,\kev). It presents
  the erratic flaring characteristic of accretion from the wind of a
  massive companion. A possible explanation for the lack of pulsations
  is that these are wiped out by the very hot Compton
  cloud. 
 
\item Finally, an absorption dip around 29\,\kev is present which, if
  interpreted as a CRSF, would identify the system unambiguously as a
  NS binary. Unfortunately, the significance is very low and it is not possible to rule out the
  possibility of calibration effects with the available data. Further
  observations are required to definitely asses this important issue.
 
\end{enumerate}

\begin{acknowledgements}
  BeppoSAX is a joint Italian-Dutch project. This research has been
  supported by research grants ESP2001-4541-PE, ESP2002-04124-C03-03
  of the Ministerio de Ciencia y Tecnolog\'{\i}a. JMT acknowledges the
  hospitality of the High Energy Astrophysics Group in T\"{u}bingen,
  Germany, under the grant Acci\'{o}n Integrada Hispano Alemana
  HA2000-0034, during which part of this work was written. IN is researcher from the Ram\'{o}n y Cajal programme of the Ministerio de Ciencia y Tecnolog\'{\i}a. This
  research has made use of the HEASARC data base at the NASA Goddard
  Space Flight Center.

\end{acknowledgements}

\end{document}